\documentclass[aps,pra,twocolumn,amsmath,amssymb,showpacs,superscriptaddress]{revtex4-1}
\usepackage{xcolor}
\usepackage{textcomp} 
\usepackage{mathrsfs,amsmath}
\usepackage{graphicx}
\usepackage{dcolumn}
\usepackage[mathlines]{lineno}
\usepackage{hyperref}
\usepackage{bm}
\usepackage{epstopdf}
\usepackage{soul}
\usepackage{epsfig}
\usepackage[normalem]{ulem}
\usepackage{bbold}
\usepackage{braket}
\usepackage{physics}
\usepackage{gensymb}
\usepackage{amsmath,amssymb}
\usepackage{booktabs}
\usepackage[table]{xcolor}
\usepackage[margin=1in]{geometry}
\usepackage{calligra}

\newcommand{\nn}{\nonumber}

\begin{document}
\title{Entanglement dynamics of light's orbital angular momentum under a Lorentz boost}
%\title{Studying the entanglement of light's orbital angular momentum under a Lorentz boost}
%\\(Entanglement, nonlocality, and the collapse of the wavefunction)}% Force line breaks with \\
%\title{Entanglement in motion}% Force line breaks with \\

\author{Moslem Mahdavifar}
\email{mahdavifar.m@gmail.com}
\email{moslem.mahdavifar@wits.ac.za}
%\affiliation{Physical Realism}
\affiliation{School of Physics, University of the Witwatersrand, Private Bag 3, Wits 2050, South Africa}

%\author{B}
%\email{}
%\affiliation{Uncertainty}
%\affiliation{School of Physics, University of the Witwatersrand, Private Bag 3, Wits 2050, South Africa}

%\author{C}
%\email{}
%\affiliation{School of Physics, University of the Witwatersrand, Private Bag 3, Wits 2050, South Africa}

%%%%%%%%abstract
\begin{abstract}
\noindent In this study, we report on the evolution of photonic orbital angular momentum (OAM) entanglement in inertial reference frames under a Lorentz boost, covering the general cases of zero and non-zero relative motion between observers of the entangled state. We find that entanglement undergoes significant changes that are observer dependent, asymptotically approaching a minimum at very large velocities close to the light cone from the viewpoint of the stationary observers in the rest frame, and degrading completely from the viewpoint of the moving observer. Our results, as demonstrated through entanglement metrics such as entanglement entropy and purity, show that OAM and OAM entanglement are observer dependent, raising pertinent questions on the invariance of such entangled states to spacetime transformations.

\end{abstract}
\maketitle
\section{Introduction}

{\large \textbf{A}} century after the development of quantum mechanics \cite{Heisenberg:1925:QUK,schrodingjsr19262, schrodinger1926undulatory}, still the theory is not  conceptually clear even to the mind of physicists. The main reason for this comes from the counterintuitive approach of quantum mechanics in explaining its physical domain. It has the underlying element of uncertainty in its foundation \cite{heisenberg1927anschaulichen}. This fundamental uncertainty produces whatever odd behavior of quantum mechanics including wave-particle duality \cite{de1924recherches} and entanglement \cite{einstein1935can,schrodinger1935discussion}. First one was observed experimentally at very early stage of the theory \cite{davisson1927diffraction}, however, the second one causes much more disputes and arguments among physics community and beyond \cite{horodecki2009quantum,gisin2009bell,bokulich2010philosophy,sep-qt-epr,esfeld2004quantum,barad2007meeting}. Test of entanglement today mostly is done through Bell's inequalities measurements \cite{bell1964einstein}. The violation of these inequalities are supposed to demonstrate the existence of this bizarre property of the quantum world which for a physical system is a nonlocal superposition of its quantum subsystems \cite{freedman1972experimental,aspect1982experimental,weihs1998violation,mahdavifar2021violating}. People have verified the violation of Bell's inequalities even with bright source of radiation that is considered to be a classical source (\cite{mclaren2015measuring} and references therein). The main element of their physical systems in achieving this result is only non-separability and to be more precise a non-separable superposition. For example, by the non-separable superposition of polarization and OAM of light \cite{allen1992orbital}. This brings us to the point where we may consider other possibilities to test the dynamics of entanglement. There have been several studies on the behavior of the entanglement in non-inertial and accelerated frames of reference \cite{alsing2003teleportation,fuentes2005alice,alsing2006entanglement,adesso2007continuous,wu2023orbital}. Most of these works show that, due to vacuum fluctuation in these frames, the structure of entanglement is degraded. On the other hand, most entanglement studies in inertial frames have concluded that entanglement is preserved for the massless particles, yet, the reported work are mixed on the entanglement dynamics of massive particles (\cite{peres2004quantum} and references therein). It is generally argued that for the highly relativistic dynamics ($v=c$), the entanglement of massive particles is not preserved in Bell tests. Yet such dynamics are physically inaccessible to massive particles. Most studies on light in inertial frames have focused primarily on polarization, which is inherently two-dimensional \cite{peres2004quantum}. Recently, we showed that for a maximally high-dimensional OAM-entangled system \cite{mair2001entanglement} generated via spontaneous parametric down-conversion (SPDC), the probability of joint detection changes under a boost applied to one of the observers’ spatial coordinates \cite{nothlawala2025orbital}. This observation was already predicted by other work, e.g., \cite{ bliokh2012spatiotemporal} for the radiation carrying OAM. We used this property to relate dispersion of OAM modes to the velocity of moving frame represented in the Lorentz contraction factor of $\gamma(v)$. The results can be summarized as a consequence of uncertainty between OAM and phase. In addition, our work did not consider the entanglement evolution and was only limited to a specific relative motion, i.e., only co-moving detection frames.

In this study, to the best of our knowledge, we consider the first principal dynamics of the OAM entanglement in moving inertial frames under a Lorentz boost. The OAM entanglement is particularly interesting because it involves a spatial mode of radiation with high-dimensional entanglement potential, in contrast to polarization, which is orientation-related and only two-dimensional. First, we study co-moving detection frames with a zero relative motion (Zero RM), and then we extend the idea to a non-zero relative motion (Non-Zero RM) between observes of the entangled state. In the case of Non-Zero RM, we distinguish between two specific viewpoints. In this scenario, we choose to have one of the observers to stay in the source frame (rest frame), then we have two possibilities for the prediction outcomes of probability amplitude. The first scheme (Non-Zero RM1) occurs in the rest frame of the source while in the second one (Non-Zero RM2) the measurements takes place from the perspective of the moving frame of reference. Our work shows that the entanglement will no longer be preserved in moving reference frames as maximally as in the case of detectors being fixed in the rest frame. We analyze the behavior of the mentioned systems by observing key quantities such as entanglement entropy, purity, effective dimensionality and negativity. These are the metrics for the entanglement dynamics in a quantum system. They undergo significant changes upon the relativistic motion of the moving frames. In both Zero RM and Non-Zero RM1 dynamics, there is a lower or upper bounds for the above metrics. Here, the degree of entanglement reduces substantially as $\gamma(v)$ increases, but entanglement still exists, though only for a few effective modes. However, the story is different from the perspective of the moving frame in the Non-Zero RM2. As will be shown, in this case the entanglement degrades completely for high values of $\gamma(v)$ regardless of modes number being entangled in the source frame. No matter how many modes are initially entangled in the source frame, the final state near the light cone (LC) becomes unentangled, i.e., the state becomes separable. 
%This poetically means:\\ ``PHOTONS DO NOT  EXPERIENCE ENTANGLEMENT."

\begin{figure*}[t!]
    \centering
    \includegraphics[width=1\textwidth]{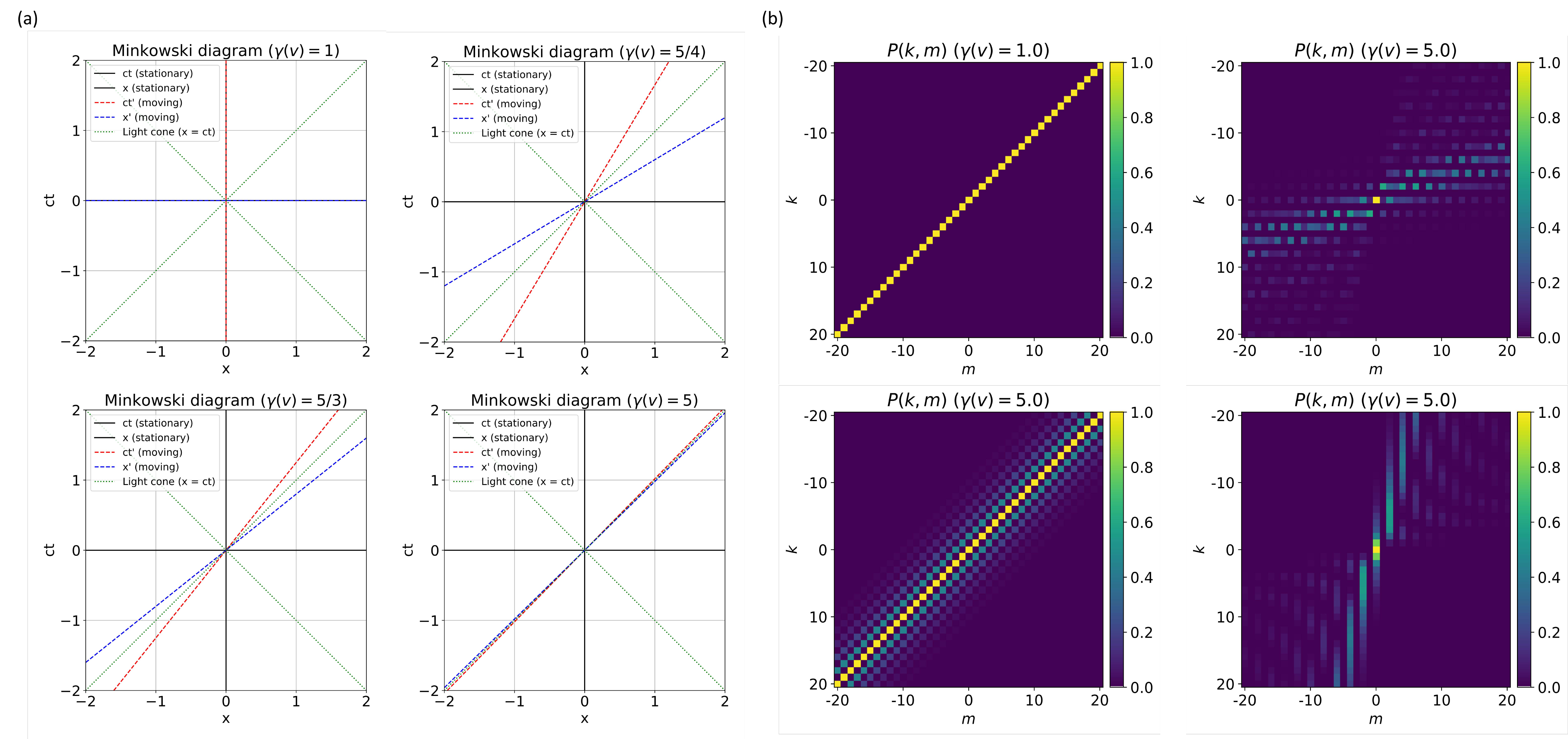}
    \caption{(a) It shows Minkowski diagrams for different relativistic velocities corresponding to $\gamma(v)=1,5/4,5/3,5$. Photon pairs that are entangled in their OAM degree of freedom are generated in the stationary reference frame of $S(ct,x)$ sent to two independent detectors, Alice (A) and Bob (B) in the moving frame of $S'(ct',x')$. The relative velocity of these two frames is $v$, and the motion is only restricted to one dimension say $x$. In the rest frame, this is seen as a length contraction in the $x$-direction of the detectors, mapping the coordinates $(x, y) \rightarrow (x/ \gamma, y)$ and $\phi \rightarrow \phi' (\phi) = \arctan (\gamma \tan(\phi))$.  The detectors project onto the OAM eigenstates $\ket{k}$ and $\ket{m}$, where Charlie (observer in $S$) observes their OAM joint detections as shown in the right column of (b) for Alice and Bob being at rest ($\gamma=1$, upper image) and moving at a relativistic speed ($\gamma(v)=5$, lower image). For $\gamma(v)=1$ the plot shows that the joint detection probability has OAM anti-correlations, i.e. non-zero detection probability for $k=-m$. However, as $\gamma(v)$ increases, the OAM dispersion appears and the width of the spectrum increases with the increment of the Lorentz factor ($\gamma(v)=5$). Right column in part (b) shows the joint detection probabilities for the Non-Zero RM1 (upper image) and Non-Zero RM2 (lower image) models and $\gamma(v)=5$.}
    \label{fig:concept}
\end{figure*}

\section{Concepts and Theory}

Here, we propose using OAM of light to find the dynamics of entanglement by exploiting spectral shifts in OAM due to a Lorentz boost. We show that this method can be used to evaluate the dynamics of related physical quantities such as entanglement entropy and purity in inertial reference frames. In such frames, length contraction rescales spatial coordinates of detectors producing measurable shifts in detected OAM correlations. For two photon states entangled in OAM, this effect modifies their joint correlations, and consequently related physical parameters.

Consider two observers of the OAM entanglement, i.e., Alice (A) and Bob (B) that are moving at relativistic velocities in  an inertial frame $S'(ct',x')$ relative to a stationary observer, here, Charlie (C), who is at the rest frame $S(ct,x)$, as depicted in FIG.~\ref{fig:concept} (a). (C) sends multiple pairs of entangled photons to (A) and (B) who act as the effective detectors of these entangled photons. The aim is to use the entangled photons to investigate the dynamics of entanglement in (A) and (B) frame via their correlated measurements. To begin, we assume that the photons are entangled in the OAM degree of freedom and are described by the two-photon state, $\ket{\Psi} = {1}/{ \sqrt{N}}( \sum_{k,m} \ket{k}_A\ket{m}_B)$, where $N$ is the normalization factor related to the number of modes contributing to the quantum state. The OAM modes $(k,m)$ should add up to zero due to angular momentum conservation. The OAM eigenstates $\ket{k} \propto \int \exp(i k \phi ) \ket{\phi} d \phi$ are characterized by an azimuth  dependent  ($\phi = \atan(y/x)$) phase profile, $ \exp(i k \phi )$, with $k$ corresponding to an OAM of $k\hbar$ per photon. Alternatively, the two-photon state can be rewritten in the azimuthal coordinate basis as $\ket{\Psi} \propto \int \ket{\phi} \ket{\phi} d \phi $, where the inner product relation $\langle \phi|\phi' \rangle = \delta(\phi' - \phi)$ holds. By defining the two-photon states in this way, the measurement on entangled state acts as a projection that maps (A)'s and (B)'s states onto wavefunctions described in terms of the $\phi$ coordinate. This means that the projection state used by Alice and Bob take the form $\langle  \psi_{A, B}  | \phi \rangle = \psi^{*}_{A(B)}(\phi)$.

The dynamics of moving frames are restricted to one dimension, e.g., $x$. Thus these frames experience a relativistic length contraction as $(x, y) \rightarrow ({x}/{\gamma}, y)$, where $\gamma(v)=(1-(v/c)^2)^{-1/2}$ is the Lorentz factor. We have two specific movements. First, both observers are co-moving or having a Zero RM. In the second one, the Non-Zero RM would be considered. This includes the dynamics of the system while one observer (A) is stationary, e.g. on the reference frame of $S$ with (C) and the other one (B) is moving away from (A) in an inertial frame. In addition, we study two distinct observer dependent measurements in the Non-Zero RM case. In the first one (Non-Zero RM1), the dynamics of the entanglement is carried out in the rest frame of the stationary observer while in the second viewpoint (Non-Zero RM2) the dynamics is carried out in the moving frame of reference. It is important to note that all the above scenarios occur in inertial frames. Moreover, the synchronization is fixed while both observers are at rest by checking joint detection and maximal entanglement observation. In the Zero RM scenario, since both observers are co-moving  their watches are synchronized. However, in the second one, there is no definite synchronization after the second observers moves away. This is due to lack of definite simultaneity in relativity.

The OAM projections that (A) and (B) perform are described using length-contracted azimuthal coordinates that are transformed as $\phi \rightarrow \phi' (\phi) = \arctan (\gamma \tan(\phi))$, resulting in the corresponding probability amplitude proportional to $\int \exp(-i(k+m) \phi'(\phi)) d \phi$. In FIG.~\ref{fig:concept} (a), the dynamics of the considered frames of reference, i.e., $S$ and $S'$ moving at different velocities corresponding to  $\gamma(v)=5/4,5/3,5$ as well as for $v = 0$ ($\gamma(v)=1$) is illustrated. Relevant joint detection probabilities to the motions in part (a) of FIG.~ \ref{fig:concept} is presented in Fig.~\ref{fig:concept} (b). However, we do not build the dynamics of entanglement from joint detection probability and instead we consider probability amplitude for the entanglement dynamics. The main reason is that the joint detection probability does not provide all the necessary information to have a complete study on the entanglement in moving frames due to its delivery based on the intensity and lacking phase information. According to the description above, three main models pertinent to the dynamics of moving frames for the joint amplitude $A(k,m)$ are considered as following, where $(k,m) \in \{-l_\text{max}, \ldots, l_\text{max}\}$ are OAM indices for the two-photon states,

\begin{enumerate}

\item \textbf{Zero RM:}
\begin{align}
A(k,m) =& \frac{1}{2\pi}\int_0^{2\pi} e^{-i (k+m) \phi'} \, d\phi \nn \\
=& \frac{1}{2\pi}\int_0^{2\pi} \frac{\gamma(v) e^{-i (k+m) \phi}}{(\gamma^{2}(v) - 1) \cos^2 \phi + 1} d\phi
\label{Z}
\end{align}

\item \textbf{Non-Zero RM1:}
\begin{equation}
A(k,m) = \frac{1}{2\pi}\int_0^{2\pi} e^{-i k \phi} e^{-i m \arctan(\gamma(v) \tan \phi)} \, d\phi
\label{NZ1}
\end{equation}

\item \textbf{Non-Zero RM2:}
\begin{equation}
A(k,m) = \frac{1}{2\pi}\int_0^{2\pi} e^{-i k \arctan(\frac{1}{\gamma(v)}{\tan \phi})}e^{-i m \phi}  d\phi
\label{NZ2}
\end{equation}

\end{enumerate}

\begin{figure*}[t!]
    \centering
    \includegraphics[width=1\textwidth]{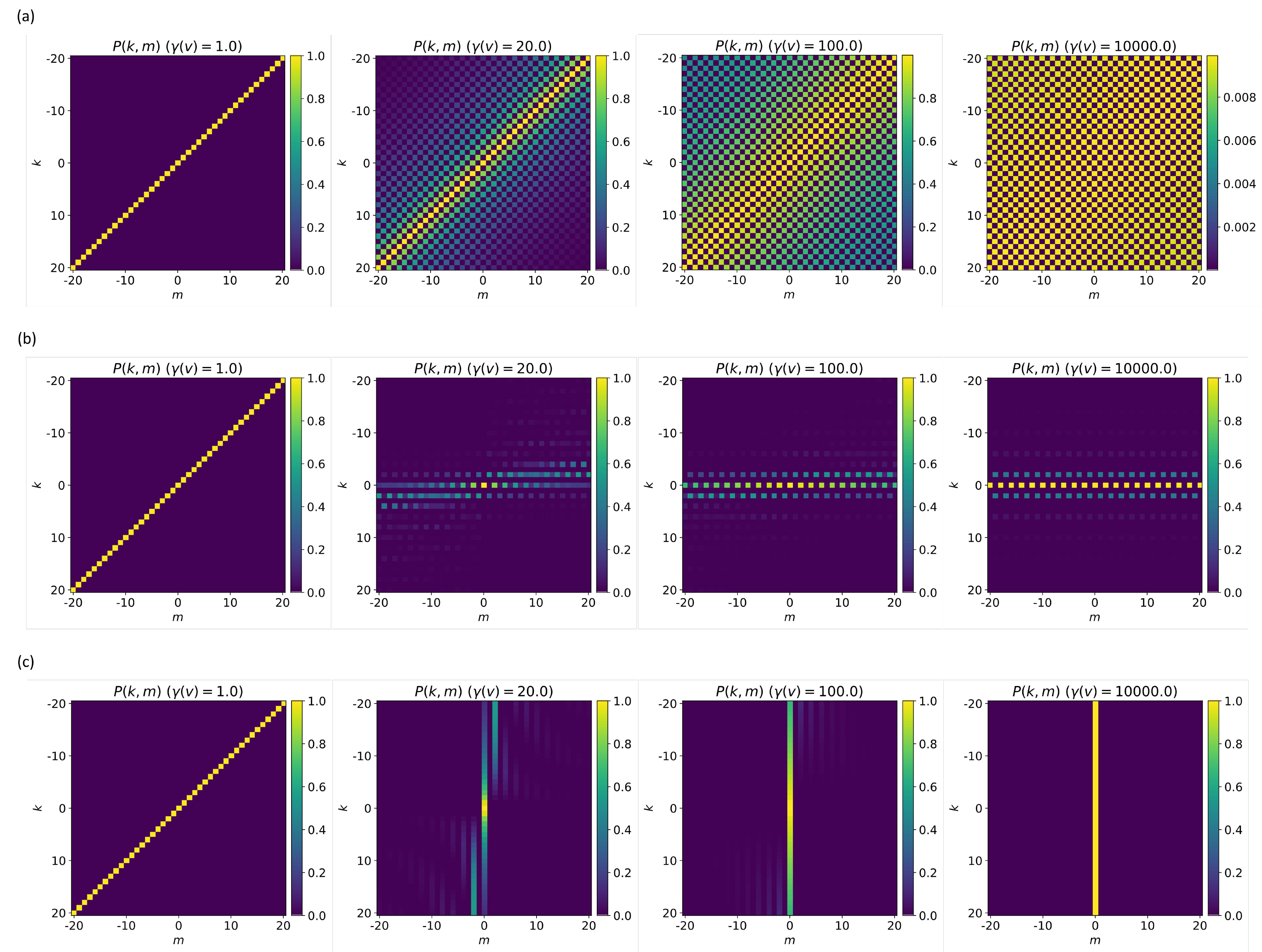}
    \caption{Joint detection probability ($P(k,m)$) is shown for $l_{max}=20$ and $\gamma(v)=1,20,100,10000$. Each row in subplots (a-c) shows one dynamical model. (a) relates to Zero RM that shows a drastic dispersion of OAM modes at high velocities. (b) displays Non-Zero RM1 in which the asymptotic behavior of entanglement is only due to three modes. (c) represents Non-Zero RM2 that includes only a single mode close to LC, hence, the state is separable and entanglement vanish entirely.}
    \label{JDP-P}
\end{figure*}

These three models represent the joint quantum probability amplitudes of detecting photons with OAM modes $k$ and $m$. First model, relates to the amplitude of observers in Zero RM. Second and third models are related to observers in Non-Zero RMs. However, we again emphasize that in the Non-Zero RM1 model, the integration is carried out by the stationary observer ($\phi$) while in the Non-Zero RM2, this happens from moving observer frame of reference ($\phi'$). The dummy integration parameters in all cases transferred to $\phi$. Also, the normalization factor of $N$ was dropped since it is not used explicitly in this study. Based on the probability amplitude $A(k,m)$, we can define a set of physical quantities related to our study as below from which we derive the dynamics of OAM entanglement in inertial frames of reference. The joint detection probability is given by the squared modulus of the amplitude matrix as
\begin{equation}
P(k,m) = |A(k,m)|^2
\label{JDP}
\end{equation}

This matrix represents the probability of detecting the two photons with OAM modes $k$ and $m$.  It is possible to find a closed analytical formula for the Zero RM case, but the Non-Zero RMs cases do not have such a closed form. Instead, they can be solved numerically and results match very well with the observed probabilities. Nevertheless, we do not need to have a closed analytical structure as a necessity in our scheme since the structure of Eqs. (\ref{Z}-\ref{NZ2}) is enough to build all the required elements in our analysis. The joint detection probability is visualized as a heatmap in the results as in the FIG.~\ref{fig:concept} (b) corresponding to the above considered cases for $\gamma(v)=1,5$ (see the caption for more details). The central point for defining physical parameters in this work is related to the definition of the state vector $\ket{\psi}$ from amplitude matrix $A(k,m)$. Also, the density matrix and reduced density matrices are defined. Given the flattened amplitude vector $\ket{\psi} = \mathrm{vec}(A)$, 
\begin{equation}
\ket{\psi} = \sum_{k,m}A(k,m)\ket{k}\otimes\ket{m}
\end{equation}
that exists in the composite Hilbert space of $\mathcal{H}_{k}\otimes\mathcal{H}_{m}$. Then, the pure bipartite state density matrix is
\begin{equation}
 \rho = \ket{\psi}\bra{\psi} = \frac{\psi \psi^\dagger}{\|\psi\|^2}   
\end{equation}
The reduced density matrices for subsystems $A$ and $B$ (corresponding to OAM modes $k$ and $m$ are obtained by partial tracing:
\begin{equation}
\rho_A = \mathrm{Tr}_B(\rho), \quad \rho_B = \mathrm{Tr}_A(\rho)  
\end{equation}
In this study, we consider a set of entanglement metrics as several measures to characterize entanglement:
\begin{itemize}

\item \textbf{Von Neumann Entropy:}
\begin{equation}
S(\rho_A) = -\mathrm{Tr}(\rho_A \log_2 \rho_A)    
\end{equation}
which quantifies the mixedness of the reduced state and entanglement.

\item \textbf{Purity:}
\begin{equation}
\mathcal{P} = \mathrm{Tr}(\rho_A^2)    
\end{equation}
measures how pure the reduced state is.

%\item \textbf{Mutual Information:}
%\begin{equation}
%I(\rho) = S(\rho_A) + S(\rho_B) - S(\rho)    
%\end{equation}    
%since the state under consideration in this study is pure, i.e., $S(\rho)=0$, hence, $I(\rho)=2S(\rho_A)=2S(\rho_B)$. As a result, the plots for the entanglement entropy and mutual %information essentially present the same information.

\item \textbf{Negativity:}
\begin{equation}
\mathcal{N}(\rho) = \sum_{\lambda_i < 0} |\lambda_i|   
\end{equation}
where \(\lambda_i\) are eigenvalues of the partial transpose of $\rho$, indicating entanglement presence.

\item \textbf{Effective Dimensionality:} Based on Schmidt coefficients $s_i$ that can be calculated from single value decomposition (SVD) of $A(k,m)$, the effective dimensionality is defined as 
\begin{equation}
 \mathcal{D}_{eff} = \frac{1}{\sum_i p_i^2} \quad \text{with} \quad p_i = \frac{s_i^2}{\sum_j s_j^2}.
 \end{equation}
 For further information about Schmidt decomposition see the appendix.
\end{itemize}
Now, by having these definitions, one can illustrate matrices as well as all metrics in this study. All mentioned operations have been done numerically that incorporates necessary math into lines of code. It produces desired physical parameters as outputs in the final forms of plots as well as a table. These illustrations have necessary and sufficient information to transfer the aim of this work which is the study of OAM entanglement evolution in moving inertial reference frames under a Lorentz boost. It includes different regime of velocities based on $\gamma(v)$. In the rest frame ($\gamma(v)=1$), one has the maximal level of entanglement, but as velocity of moving frames increase the above metrics start to develop changes. At the high velocities close to LC, changes are significant. We discuss consequences of this evolution in details in the sections \ref{R} and \ref{dis}.

\begin{figure*}[t!]
    \centering
    \includegraphics[width=1\textwidth]{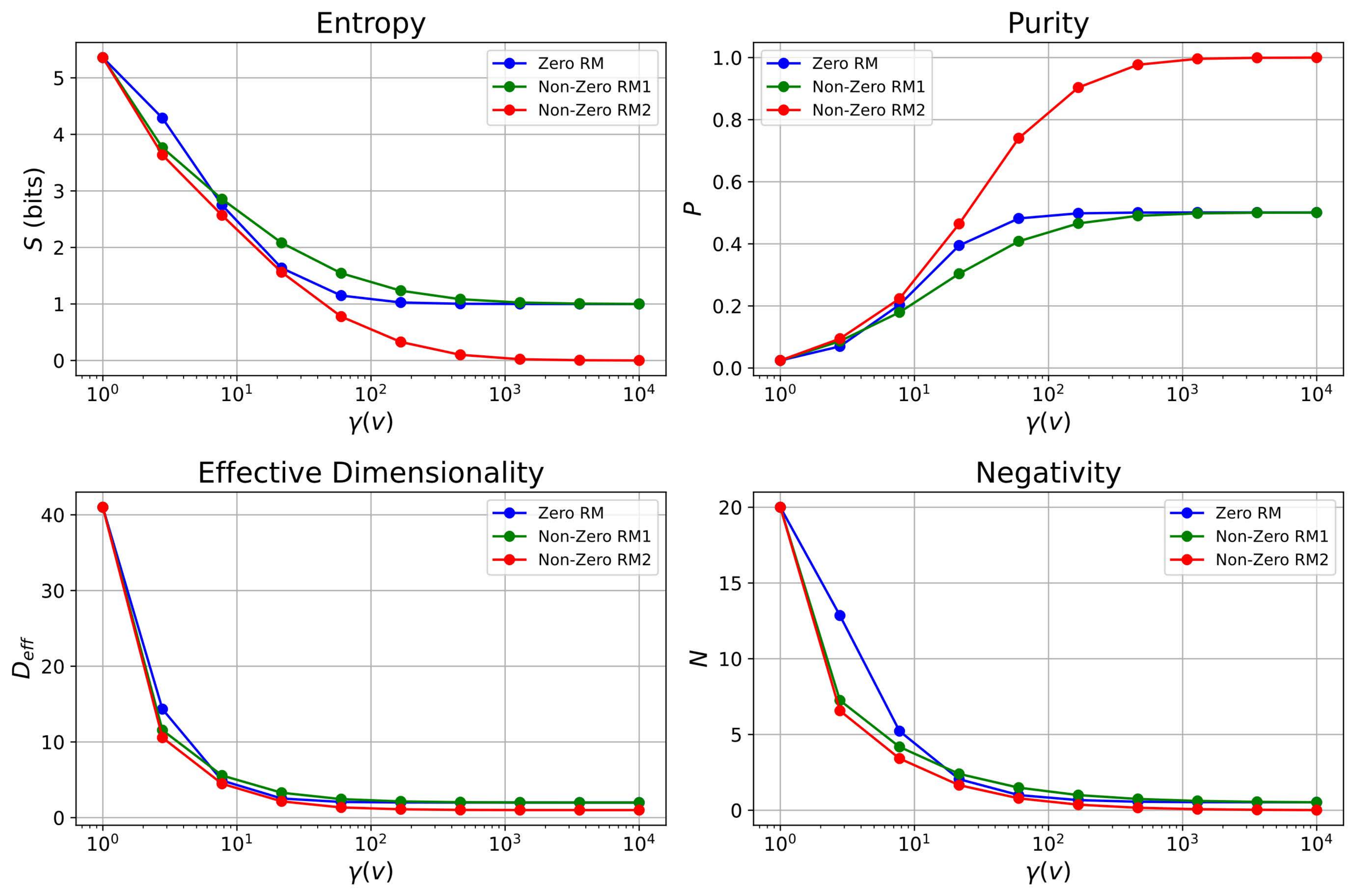}
    \caption{Metrics for $l_{max}=20$ and $\gamma(v)=1-10000$ in $\text{log}_{10}$ basis. From the stationary observer's viewpoint the quantum state will remain entangled though reach a minimum degree asymptotically at the LC. However, from the viewpoint of the moving frame's observe, the entanglement will be vanished completely at the LC. Therefore, for this observer the state is separable.}
    \label{metrics}
\end{figure*}

\renewcommand{\arraystretch}{1.3}
\rowcolors{2}{gray!10}{white}

\begin{table*}[t!]

\renewcommand{\arraystretch}{1.5} % Optional: adds row spacing for readability
\begin{tabular}{|c|l|l|l|l|}
\hline
\textbf{Physical variables} & \textbf{Metrics} & \textbf{Zero RM} & \textbf{Non-Zero RM1} & \textbf{Non-Zero RM2}\\
\hline

% Subdomain 1
            & $\mathcal{S}(bits)$  & 5.3576  & 5.3576  & 5.3576 \\
            & $\mathcal{P}$  & 0.0244  & 0.0244  & 0.0244 \\
$\gamma(v)=1,~l_{max}=20$       & $\mathcal{MI}(bits)$  & 10.7151  &  10.7151  & 10.7151  \\
            & $\mathcal{N}$ & 20.0000 & 20.0000 & 20.0000 \\
            & $\mathcal{D}_{\text{eff}}$ & 41.00 & 41.00 & 41.00 \\
\hline

% Subdomain 2
            & $\mathcal{S}$ & 3.3717 & 3.2377  & 3.0279 \\
            & $\mathcal{P}$ & 0.1344 & 0.1337  & 0.1563 \\
$\gamma(v)=5,~l_{max}=20$       & $\mathcal{MI}$ & 6.7434 & 6.4754  & 6.0558 \\
            & $\mathcal{N}$ & 7.8350 & 5.3251  & 4.5732 \\
            & $\mathcal{D}_{\text{eff}}$ & 7.44 & 7.48  & 6.40 \\
\hline

% Subdomain 3
            & $\mathcal{S}$ & 1.6948 & 2.1295  & 1.6314 \\
            & $\mathcal{P}$ & 0.3833 & 0.2946  & 0.4431 \\
$\gamma(v)=20,~l_{max}=20$       & $\mathcal{MI}$ & 3.3896 & 2.4934  & 1.7478 \\
            & $\mathcal{N}$ & 2.1838 & 2.4934  & 1.7478 \\
            & $\mathcal{D}_{\text{eff}}$ & 2.61 & 3.39  & 2.26 \\
\hline

% Subdomain 4
            & $\mathcal{S}$ & 1.0644 & 1.3631  & 0.5106 \\
            & $\mathcal{P}$ & 0.4938 & 0.4428  & 0.8370 \\
$\gamma(v)=100,~l_{max}=20$       & $\mathcal{MI}$ & 2.1289 & 2.7262  & 1.0212 \\
            & $\mathcal{N}$ & 0.7896 & 1.2057  & 0.5349 \\
            & $\mathcal{D}_{\text{eff}}$ & 2.03 & 2.26  & 1.19 \\
\hline

% Subdomain 5
            & $\mathcal{S}$ & 1.0189 & 1.1978  & 0.2755 \\
            & $\mathcal{P}$ & 0.4992 & 0.4723  & 0.9219 \\
$\gamma(v)=200,~l_{max}=20$       & $\mathcal{MI}$ & 2.0378 & 2.3956  & 0.5511 \\
            & $\mathcal{N}$ & 0.6403 & 0.9396  & 0.3219 \\
            & $\mathcal{D}_{\text{eff}}$ & 2.00 & 2.12  & 1.08 \\
\hline

% Subdomain 6
            & $\mathcal{S}$ & 0.9997 & 1.0147  & 0.0104 \\
            & $\mathcal{P}$ & 0.5011 & 0.4993  & 0.9983 \\
$\gamma(v)=2000,~l_{max}=20$       & $\mathcal{MI}$ & 1.9994 & 2.0295  & 0.0208 \\
            & $\mathcal{N}$ & 0.5292 & 0.5841  & 0.0433 \\
            & $\mathcal{D}_{\text{eff}}$ & 2.00 & 2.00  & 1.00 \\
\hline

% Subdomain 7
            & $\mathcal{S}$ & 0.9995 & 1.0004  & 0.0008 \\
            & $\mathcal{P}$ & 0.5011 & 0.5008  & 0.9999 \\
$\gamma(v)=10000,~l_{max}=20$       & $\mathcal{MI}$ & 1.9990 & 2.0008  & 0.0017 \\
            & $\mathcal{N}$ & 0.5270 & 0.5246  & 0.0096 \\
            & $\mathcal{D}_{\text{eff}}$ & 2.00 & 2.00  & 1.00 \\
\hline

\end{tabular}
\caption{It includes different values of metrics (entanglement entropy ($\mathcal{S}$), purity($\mathcal{P}$), mutual information($\mathcal{MI}$), negativity($\mathcal{N}$), and effective dimensionality ($\mathcal{D}_{\text{eff}}$)) across $\gamma(v)=1,5,20,100,200,2000,10000$ and $l_{max}=20$. While the metrics at rest have the maxima except for the purity with a minima, the effect of motion takes all to their extremum at $\gamma(v)\gg1$.}
\label{table}
\end{table*}

\section{Proposed Experiments}
To emulate these predicted relativistic effects on the evolution of entanglement, one can use photons generated from spontaneous parametric down conversion (SPDC) similar to \cite{nothlawala2025orbital}. However, in that case, the analysis can include only joint detection probabilities and not from the reconstruction of the state through quantum state tomography (QST). For the joint detection, two-photon states are imaged onto independent spatial light modulators (SLMs) with holograms simulating the distorted detection modes (based on Eqs. (\ref{Z}-\ref{NZ2})). The resulting modulated photons are collected with single-mode fibers and detected in coincidence counts. The experimentally measured joint probability spectra can then be compared with theoretical predictions. Our Zero RM results on the joint detection matches very well with \cite{nothlawala2025orbital}. It is worth mentioning that, in this approach, the SLM resolution (pixel size) is important due to the rapid phase oscillation in Eq. (\ref{NZ2}). Another effective approach to determine the dimensionality and the purity of entanglement experimentally is to use the method implemented in \cite{nape2021measuring} which in contrast to QST is timely efficient.
\begin{figure*}[t!]
    \centering
    \includegraphics[width=1\textwidth]{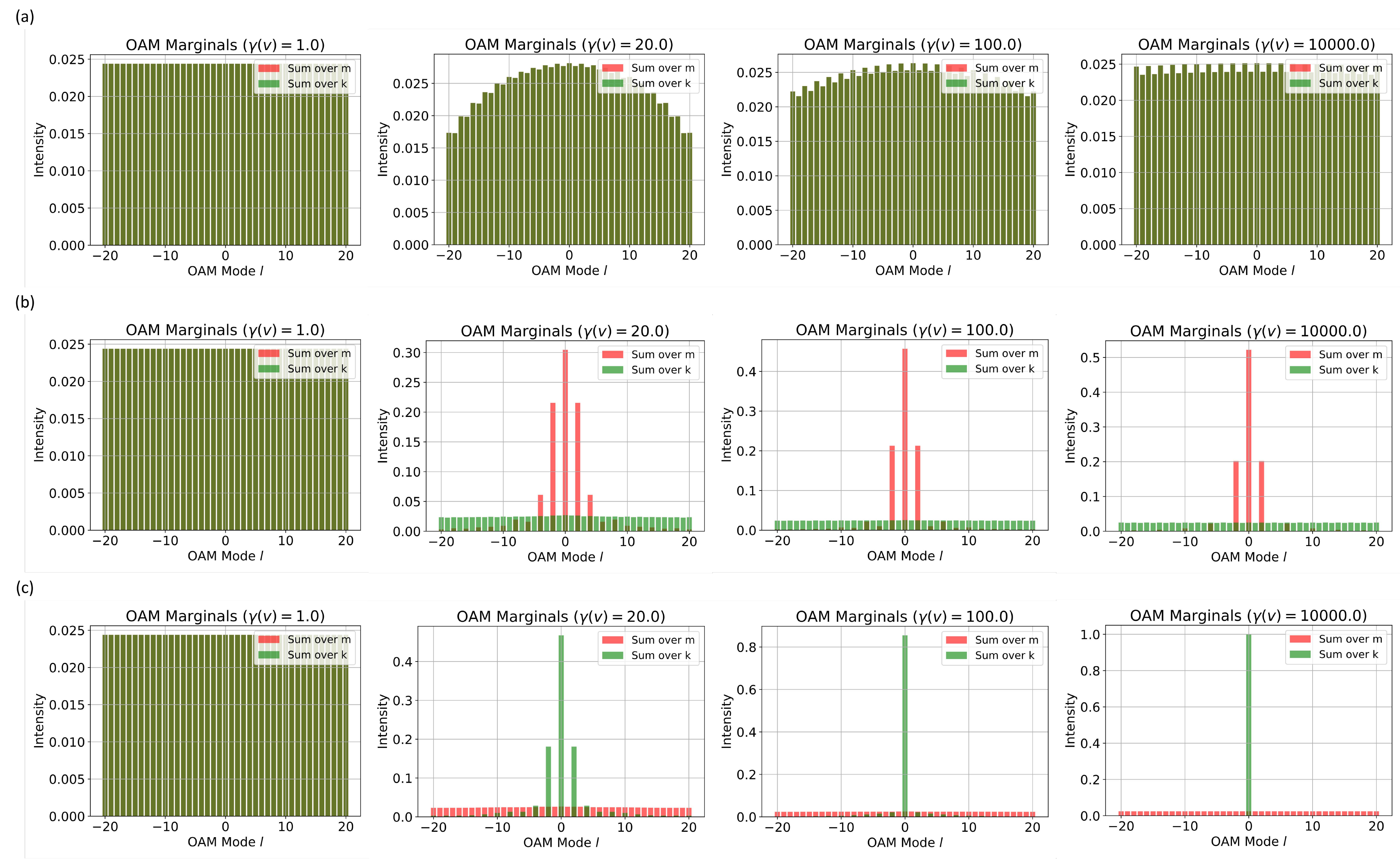}
    \caption{OAM marginals for $l_{max}=20$ and $\gamma(v)=1,20,100,10000$. (a) In the Zero RM, the information regarding the entanglement is scrambled in a way that no longer one can find the obvious track of its present. This is due to the extreme OAM dispersion. However, in the Non-Zero RMs, the entanglement behavior is clear. (b) displays the dynamics of entanglement in the Non-Zero RM1 that asymptotically approaches a minimum degree close to LC. Here, mainly five modes contribute to the joint detection. (c) illustrates the entanglement evolution in the Non-Zero RM2 that at LC will perfectly vanish.}
    \label{marginals}
\end{figure*}
\section{Results}\label{R}
\begin{figure*}[t!]
    \centering
    \includegraphics[width=1\textwidth]{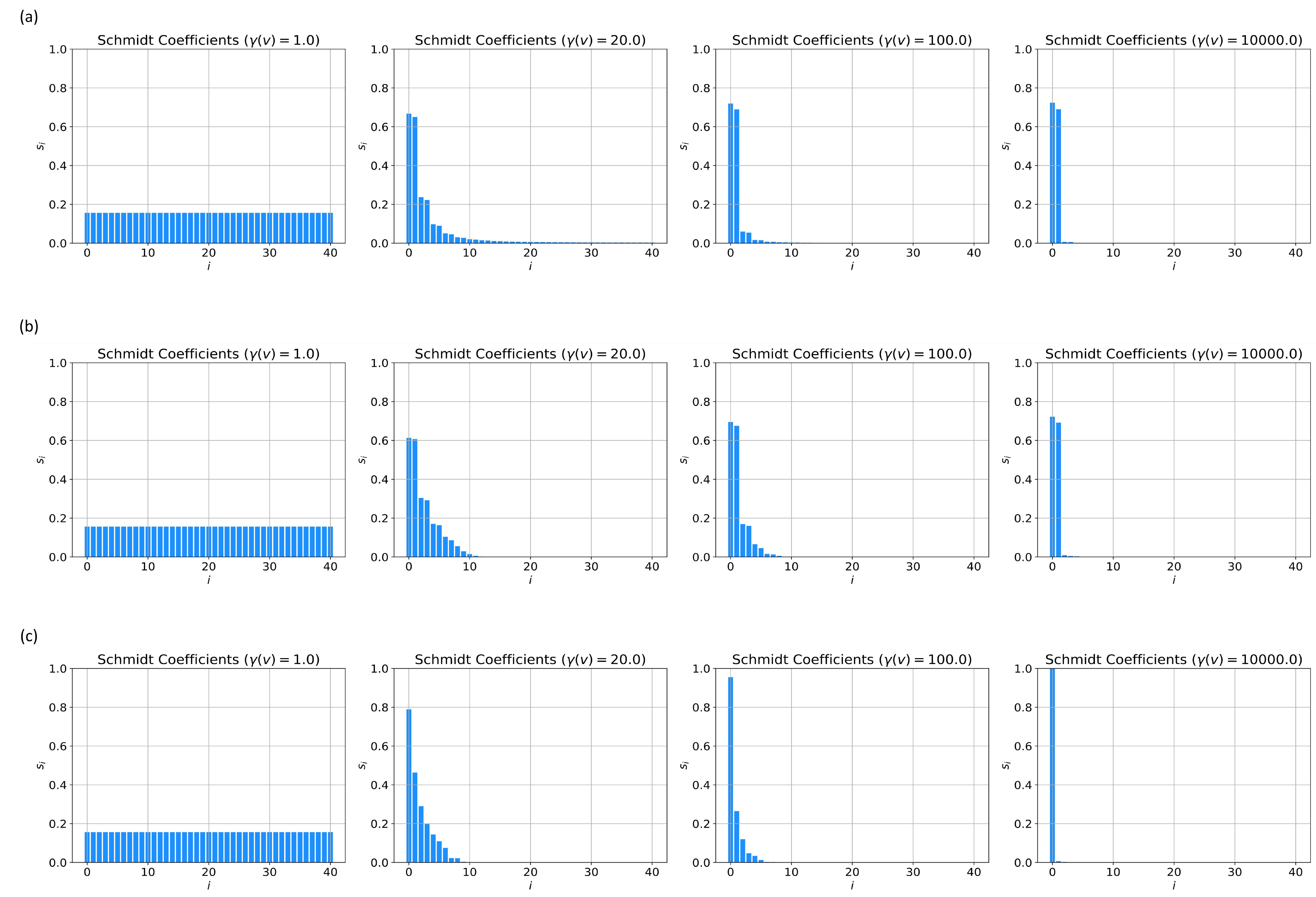}
    \caption{Schmidt decomposition coefficients related to the OAM entanglement evolution as a measure for the entanglement dynamics related to $\gamma(v)=1,20,100,10000$. Close to the LC, the Zero RM (a), and the Non-Zero RM1 (b) only have two effective modes. However, for the Non-Zero RM2 (c), the number of effective mode is only one asymptotically. This confirms the absence of entanglement in the case of Non-Zero RM2 model at LC.}
    \label{Schmidt}
\end{figure*}

Now, let's take a closer look at the data from the above concepts regarding the entanglement dynamics. It is worth mentioning that we fixed the maximum value of OAM as $l_{\text{max}}=20$ in this study, therefore, in all plots OAM spectra is $(k,m) \in \{-20, \ldots, 20\}$. In FIG.~ \ref{JDP-P}, the joint detection probability is illustrated that is based on Eq. (\ref{JDP}). It includes three models in three rows (a-c) for $\gamma(v)=1,20,100,10000$. The first row shows Zero RM, second and the third ones are related to Non-Zero RM1 and Non-Zero RM2, respectively. The joint probability distribution is necessary information for testing the predictions of this study. It can be obtained via coincidence counts in a quantum setup that produces maximally entangled states, e.g., from SPDC. In FIG.~\ref{metrics}, all four metrics are presented. It displays metrics for the range of $\gamma(v)=1-10000$. The plot shows four subplots covering the dynamics of four metrics as entanglement entropy ($\mathcal{S}$), purity ($\mathcal{P}$), negativity $\mathcal{N}$, and effective dimensionality $\mathcal{D}_{\text{eff}}$. For example, for the models related to the perspective of the stationary observer(s) at $S$ (Zero RM and Non-Zero RM1), the $\mathcal{S}$ approaches 1 and the $\mathcal{P}$ approaches 0.5. These results indicate the lowest possible degree of entanglement from the perspective of the observers in the rest frame of $S$. In high $\gamma(v)$ regime ($\gamma(v) \gg 1$), all the modes contribute to the Zero RM model, however, the OAM dispersion here is extreme (see FIG.~\ref{JDP-P} (a) and FIG.~\ref{marginals} (a)). In the Non-Zero RM1, mainly a few modes contribute to the entanglement with the main single-detection contribution from $l=0$ and symmetric intensities for other modes (see FIG.~\ref{JDP-P} (b) and FIG.~\ref{marginals} (b)). However, as stated earlier the Non-Zero RM2 model has the most striking behavior and shows the complete degradation of entanglement. In this case, $\mathcal{S}$ and $\mathcal{N}$ approach 0, $\mathcal{P}$ and $\mathcal{D}_{\text{eff}}$ approach 1, asymptotically. These values are the main features of a separable state and confirm the lack of entanglement for the physical system under consideration. Mentioned asymptotic behavior of the metrics  for all models can also be inferred from the dynamical matrices in FIG.~\ref{JDP-P} (a-c) as well. Part (c) clearly shows that the only modes contributing to the state at $\gamma(v)=10000$ comes from $\ket{00}$ state which is a separable state (see also FIG.~\ref{marginals} (c)). The main mathematical reason for this singular feature originates from this fact that the Schmidt decomposition at this asymptotic limit has only one mode contribution (FIG.~\ref{Schmidt} (c)) and the cumulative Schmidt probability ($\sum_{i} p_i$) is satisfied with only a single mode. As a matter of fact, our numerical results is closely connected to this math. In the appendix, a detailed record of these elements is presented for all models in FIG.~\ref{Schmidt}. This is obviously manifested in the effective dimensionality metric $\mathcal{D}_{\text{eff}}=1$ in FIG.~\ref{metrics} as well as in the Table~\ref{table}. These asymptotic behaviors of the OAM entanglement is analogous to the behavior of the gravitational field of a black hole from the perspectives of a far-away stationary observer and falling observer through the horizon. For the far observer, the falling observer would never reach the horizon. On the other hand, the falling observer does not observe anything strange and can pass the horizon in a proper time (see, e.g., \cite{d1992introducing}). For the dynamics of the entanglement under this study, the rest frame observer(s) would never observe the separability of the state though the number of entangled modes decreases to a lower limit, but still entanglement exist to some degree. However, from the perspective of the observer close to the LC, the entanglement does not exist and the state is separable. This can be derived from Eq. (\ref{NZ2}) under this condition that close to LC, $\gamma(v)$ becomes very large and, therefore, the only effective mode contributing to the state is $\ket{00}$ with the probability equal to unity. Moreover, one can clearly obtain the modes contribution to the final state by considering the OAM marginals, i.e., the OAM modes contribution to the joint detection probability as,
\begin{equation}
    P(k)=\sum_{m}P(k,m)
\end{equation}
FIG.~\ref{marginals} shows this for $\gamma(v)=1,20,100,10000$. Furthermore, the detailed values of the above metrics are provided in Table~\ref{table}.

Based on the results of this study, one can make a prediction regarding the photon's OAM. The final state contribution at LC comes only from $\delta_{m0}$ (see Eq. (\ref{NZ2})), regardless of the OAM value of the pump source. This indicates that, for an observer approaching the LC, OAM becomes irrelevant. This, in turn, calls into question the common assumption that photons fundamentally carry OAM. Consequently, it raises the critique that OAM and OAM entanglement in photons are likely observer-dependent. Surely, OAM conservation exists, and the non-zero OAM of pump source is transferred to the OAM mode of detected photon in the rest frame $S$. This prediction can be tested using joint detection probability for the $\gamma(v)\gg1$ and with non-zero OAM value of the pump source.

%\begin{figure*}[t!]
 %   \centering
  %  \includegraphics[width=1\textwidth]{Figures/cumulative1.pdf}
   % \caption{Cumulative probability of the Schmidt decomposition related to the OAM entanglement evolution as a measure for the entanglement related to $\gamma(v)=1,20,100,10000$. At high $\gamma(v)$ values there are only two effective modes that contribute to the spectra of (a), the Zero RM, and (b), the Non-Zero RM1. However, in the Non-Zero RM2 (c), only one mode is dominant indicating separability of the state in this limit.}
  %  \label{cumulative}
%\end{figure*}
\section{Discussion}\label{dis}
In this section, we emphasize the conceptual implications of our findings on the dynamics of OAM entanglement in inertial frames. According to quantum mechanics, entangled states are inherently nonlocal, a counterintuitive consequence of the spontaneous collapse of the wavefunction. The prevailing view holds that, regardless of the spatial separation between observers, state reduction occurs instantaneously upon measurement. The physics community has long sought to test this notion, most notably through Bell inequality experiments based on the assumption of hidden variables representing pre-measurement realities. However, as stated earlier nonseparability alone is sufficient to violate such inequalities. Our results indicate that existing viewpoints of entanglement are only effective descriptions. The absence of definite physical realism in quantum mechanics likely stems from the theory’s incomplete account of its physical domain across all dynamical regimes.

We further demonstrate that OAM entanglement is observer-dependent. This study extends the concepts of nonlocality and non-separability of OAM entanglement to the extreme physical limit, i.e., near the LC. Close to the LC, only the separable part of the state remains observable from the perspective of a moving observer, confirming that non-separability cannot exist on the LC.

It is known that the underlying physics of entanglement is based upon the uncertainty principle \cite{einstein1935can}. In quantum field theory (QFT) \cite{weinberg1995quantum}, the field and its conjugate exhibit zero uncertainty outside the LC. Our findings refine this understanding by showing that the uncertainty on the LC itself also vanishes. These results suggest that the fundamental root of the uncertainty principle in quantum mechanics is relativity, the relative motion between the observer and the physical system gives rise to uncertainty. The uncertainty does not hold on the LC nor the entanglement.\\
\section{Conclusion}
In conclusion, we have studied the OAM entanglement dynamics in inertial moving frames of reference under a Lorentz boost. In contrast to the common view that entanglement is a nonlocal property in quantum mechanics, our findings demonstrate that OAM entanglement is observer dependent. This result provides further insight by shedding light on the uncertainty in quantum mechanics. It refines the validity limit of the uncertainty principle, revealing a more compelling cause for its existence in the quantum world, namely, the relative motion between observers and the quantum system. We again emphasize that it is relativity itself that gives rise to uncertainty in quantum mechanics. This uncertainty disappears when the observer and the quantum system share the same dynamics. Finally, based on our study, we predict that there is no OAM relevance for an observer approaching the LC. These findings are raising pertinent questions on the invariance of OAM and OAM entanglement under spacetime transformations.
\section*{Appendix}\label{appendix}
Here, we present the Schmidt decomposition corresponding to the considered OAM spectra and the associated $\gamma(v)$ values. The components of the Schmidt decomposition are shown in Fig.~\ref{Schmidt}. The subplots follow the same ordering as in the preceding figures: the first row corresponds to the Zero RM case, while the second and third rows correspond to the Non-Zero RM1 and Non-Zero RM2 cases, respectively. These plots provide consistent information regarding the behavior of the relevant metrics and, consequently, the evolution of OAM entanglement. 

As $\gamma(v)$ increases, the number of contributing Schmidt modes decreases. At velocities approaching the LC, only two effective Schmidt modes remain for the Zero RM and Non-Zero RM1 cases. In contrast, for the Non-Zero RM2 case, only a single mode persists at the LC. Thus, from the perspective of inertial reference frames in motion, no entanglement survives at the LC.

\section*{Acknowledgments}
The author expresses gratitude to Andrew Forbes and Bertus Jordan for their valuable discussions on aspects of this work.

\bibliographystyle{unsrt}
\bibliography{references.bib}

%\appendix

\end{document}